\documentclass[12pt]{spieman}  
\usepackage{amsmath,amsfonts,amssymb}
\usepackage{graphicx}
\usepackage{float}
\usepackage{setspace}
\usepackage{tocloft}

\title{Enabling Continuous Operations for UAVs with an Autonomous Service Network Infrastructure}

\author[a]{Michael Rosenberg}
\author[b]{John Henry Burns}
\author[a]{Deeraj Nagothu}
\author[a,*]{Yu Chen}

\affil[a]{Dept. of Electrical \& Computer Engineering, Binghamton University, Binghamton, NY 13902}
\affil[b]{Dept. of Computer Science, Binghamton University, Binghamton, NY 13902}

\cftpagenumbersoff{figure}
\cftpagenumbersoff{table} 

\begin{document} 
\maketitle

\begin{abstract}
One of the major restrictions on the practical applications of unmanned aerial vehicles (UAV) is their incomplete self-sufficiency, which makes continuous operations infeasible without human oversights. The more oversight UAVs require, the less likely they are going to be commercially advantageous when compared to their alternatives. As an autonomous system, how much human interaction is needed to function is one of the best indicators evaluating the limitations and inefficiencies of the UAVs. Popular UAV related research areas, such as path planning and computer vision, have enabled substantial advances in the ability of drones to act on their own. This research is dedicated to in-flight operations, in which there is not much reported effort to tackle the problem from the aspect of the supportive infrastructure. In this paper, an Autonomous Service network infrastructure (AutoServe) is proposed. Aiming at increasing the future autonomy of UAVs, the AutoServe system includes a service-oriented landing platform and a customized communication protocol. This supportive AutoServe infrastructure will autonomize many tasks currently done manually by human operators, such as battery replacement. A proof-of-concept prototype has been built and the simulation experimental study validated the design. 
\end{abstract}


{\noindent \footnotesize\textbf{*}Corresponding Author: Yu Chen,  \linkable{ychen@binghamton.edu} }





\section{Introduction}

The past decade is witnessing the proliferation of unmanned aerial vehicles (UAV), also known as drones, in the era of the Internet of Things (IoT) \cite{chen2017enabling, chen2016smart}. One of the major restrictions on the practical applications of UAVs is the low-level self-sufficiency. The more oversight UAVs require, the less likely they are going to be commercially advantageous when compared to their alternatives. These systems are capable of navigating and exploring complex environments on their own. However, when one considers the required human interaction, many of these advances lose their pragmatic commercial applications. Therefore, a new system that is able to perform tasks typically done by humans is required to enable UAVs be \emph{fully} autonomous. Similar to larger commercial aircraft, if the Aerial Platform (AP) is sitting idle its' ability to work efficiently diminishes. In addition, it is highly desired to minimize this ``downtime'' and maximize the efficiency of each individual UAV by providing a network of landing platform (LP)'s supported with a customized communication protocol.

There is very little research into any type of infrastructure for autonomous vehicles \cite{gopalswamy2018infrastructure, chen2016operations}. However, it is a rich domain of applied research that could have significant implications on the practicality of UAVs. The most common complaint about UAVs is their short flight time. This means that even if they are operating fully on their own in flight, their mission will most likely be less than 45 minutes. Given the current state of available UAV battery technology, we found automated battery exchange to be a natural starting point.

Due to the currently limited flight time, UAVs are impractical in applications where the system is supposed to operate for extended periods of time \cite{gong2018flight}. If they are required to be used in such a circumstance, a human will most likely be required to keep the drones powered and in good condition. This creates a serious limitation on their adoption into industry. With an automated process for battery swapping and charging, a system's flight time is no longer an upper bound for the system's mission time and the system can manage more complex tasks without human interaction at each step.

Directly related to a maximum flight time is maximum flight distance \cite{scherer2016persistent}. UAVs are severely limited by the distance they can fly on their own given current power storage limitations. Using one of these service platforms does not fix this problem directly, but rather takes it into account to maintain an efficient system. The UAV will still only be able to fly the maximum flight distance from the platform, which is likely not far enough for most applications. However, by using a distributed network of these platforms, a UAV could travel across the network, stopping when it requires a new battery or other maintenance. Again, one can remove in flight restrictions with a supportive infrastructure. 

This work introduces a distributed network of \emph{AutoServe} stations that would be capable of providing the UAVs with any needed maintenance such as battery and cargo swapping. Most of the maintenance operations are traditionally performed manually by human operators instead of autonomously. This is a major bottleneck on any UAV-based system, and as the system scales more and more human oversight would be required to keep the system operational. By automating as much of the human tasks as possible, it allows the system to get closer to real autonomy and therefore much higher efficiency than we have seen thus far.

The rest of the paper is organized as follows. Section \ref{sec:rel} gives a brief overview of the related work. Section \ref{sec:sys} introduces the proposed AutoServe architecture, and Section \ref{sec:comm} details the design rationale of the communication protocol. Section \ref{sec:exp} presents the results of our experimental study. Section \ref{sec:conclu} concludes this paper with discussions on the future work.


\section{Related Work}
\label{sec:rel}

While UAVs have gained a lot of attention, there is very little research into designing a supportive infrastructure for them. Most research that aims at increasing their independence tackles the problem from the aspect of flight such as navigation, control, and vision. While these efforts are crucial, they only address issues related to flight. We hope this paper expands the current research of UAV autonomy into a new territory, focusing on the crucial moments between flights.

\subsection{Flight Path Optimization}

Flight path planning is an active research field in the realm of UAVs \cite{challita2018deep, lin2009uav, yoon2017adaptive}, and integrates very naturally with the AutoServe system presented by this paper. The flight path planning consists of two main technologies: collision avoidance and route optimization with obstacle avoidance. Collision avoidance is the study of using \emph{air-data} and sensors, which usually include LIDAR (Light Detection and Ranging), sonar, or camera, to indicate the position of other vehicles in the vicinity \cite{lin2017sampling, zhang2018collision}. Route optimization with obstacle avoidance is the ability to take two or more points, and develop a path between them that achieves a certain level of predefined efficiency, and avoids collision with any obstacles such as trees or traffic signs \cite{baek2019optimal, liu2019real}.
        
Since the overall goal of the proposed AutoServe system is to increase the flight distance, route optimization is very important. It is suggested that each UAV makes use of sensors to capture the environment into an understandable form \cite{zhang2011online}. The data represents possible obstructions in the UAVs route between two points, as well as different paths to take, allowing the UAV to converge on the optimal solution. 

 \begin{figure}[hbt]
    \centering
    \includegraphics[width=0.5\linewidth]{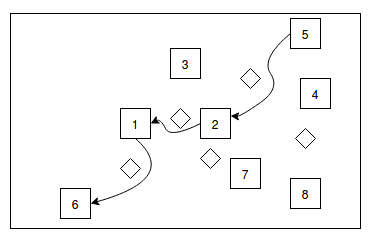}
    \caption{Example of Route Optimization with LP Graph Traversal.}
    \label{fig:path_planning}
 \end{figure}

Our AutoServe network can be considered as a graph with each LP as a node and the path between two LPs as an edge, graph traversal solutions are applicable to the trips between LPs. In Fig. \ref{fig:path_planning}, the LPs are represented by numbered squares, and the diamonds represent obstructions. The theoretical route taken by a UAV that intends to travel from LP \#5 to LP \#6 is illustrated in Fig. \ref{fig:path_planning}. Instead of a standard implementation of Dijkstra's algorithm \cite{jasika2012dijkstra}, pairing a route optimization or path planning algorithm with a customized implementation of Dijkstra's algorithm could produce an efficient flight path between many nodes. At each stage, only the LPs (or vertices) that are within a predefined distance can be considered. Another difference is that this implementation would favor the LPs that are furthest away but still within a safe range. 
        
\subsection{MAVLink Protocol}

\begin{figure}[htb]
            \centering
            \includegraphics[width=0.8\linewidth]{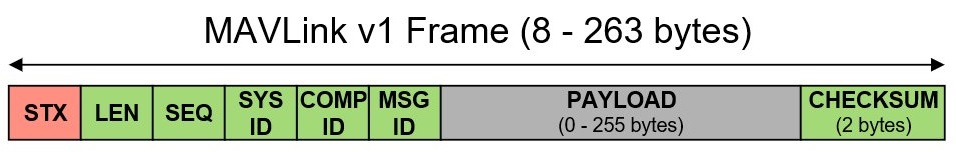}
            \caption{MAVLink v1 Packet Structure \cite{koubaa2019micro}.}
            \label{fig:my_label_2}
        \end{figure}

In order to scale such an architecture, strict protocols and clear lines of communication must be established to maintain a functioning network. It requires a central authority where all events are stored for planning. It also needs a certain level of distributed planing control in order to handle anomalies on its own, such as two UAVs wanting to use the same platform. The messaging protocol we chose was MAVLink \cite{atoev2017data}. Firstly released in 2009, MAVLink is a highly adopted, lightweight protocol to send messages between drones and control stations \cite{marty2013vulnerability}. The protocol contains many useful features such as support for multiple connections (up to 255 concurrent), programming languages (Python, C/C++, Java, etc.) and system architectures (ARM, Linux, macOS, Windows, etc.), which make MAVLink quite desirable \cite{mavlink2020mavlink}. Figure \ref{fig:my_label_2} shows the structure and fields of the MavLink V1 frames.

\begin{figure}[H]
            \centering
            \includegraphics[width=0.8\linewidth]{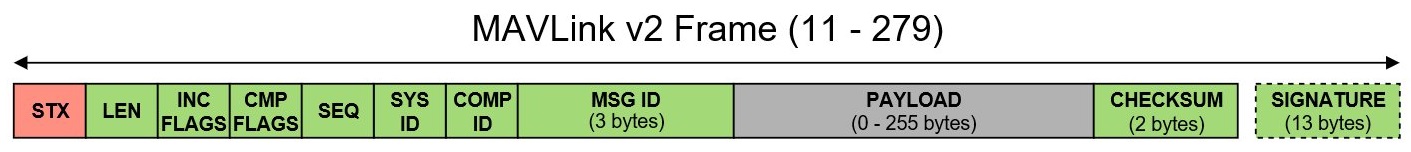}
            \caption{MAVLink v2 Packet Structure \cite{koubaa2019micro}.}
            \label{fig:my_label_3}
        \end{figure}

Due to the mission-critical nature of UAV Operations, security is among the top concerns when developing these systems. The lightweight nature of the protocol also means encryption is not implemented at this level, thus exposing the contents of messages when transmitted on an insecure channel. Earlier versions of the protocol also lacked a message signing feature, opening the door for a multitude of attacks such as Man-in-the-Middle (MITM), Eavesdropping and Denial of Service (DoS) \cite{butcher2013securing}. When version 2 of the protocol was released in 2017, as shown by Fig. \ref{fig:my_label_3}, a new signature field was added to each message that is generated using the packet, timestamp and a shared secret key that is stored on both ends of the link \cite{mavlink2020mavlink2}. This addition allows operators to verify that communications have been sent by the source they claim to be. For this reason, MAVLink v2 was adopted in our design.

        \begin{figure}[H]
            \centering
            \includegraphics[width=0.8\linewidth]{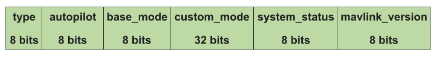}
            \caption{An example of MAVLink heartbeat message \cite{koubaa2019micro}.}
            \label{fig:mavlink_heartbeat}
        \end{figure}

An important aspect of the MAVLink protocol is its heartbeat protocol, which is used to determine the presence of a system \cite{koubaa2019micro}. Figure \ref{fig:mavlink_heartbeat} presents an example of the MAVLink heartbeat message. It also is used to express system attributes such as vehicle type, flight stack, system id, component type, and flight mode. A predetermined number of heartbeat messages within a predefined length of time defines whether a device is connected to the network or not. Certain attributes could be added to this protocol that are necessary to support our AutoServe's particular requirements. 

\section{System Architecture}
\label{sec:sys}

\subsection{System Overview}

\begin{figure}[t]
  \centering
  \includegraphics[width=\linewidth]{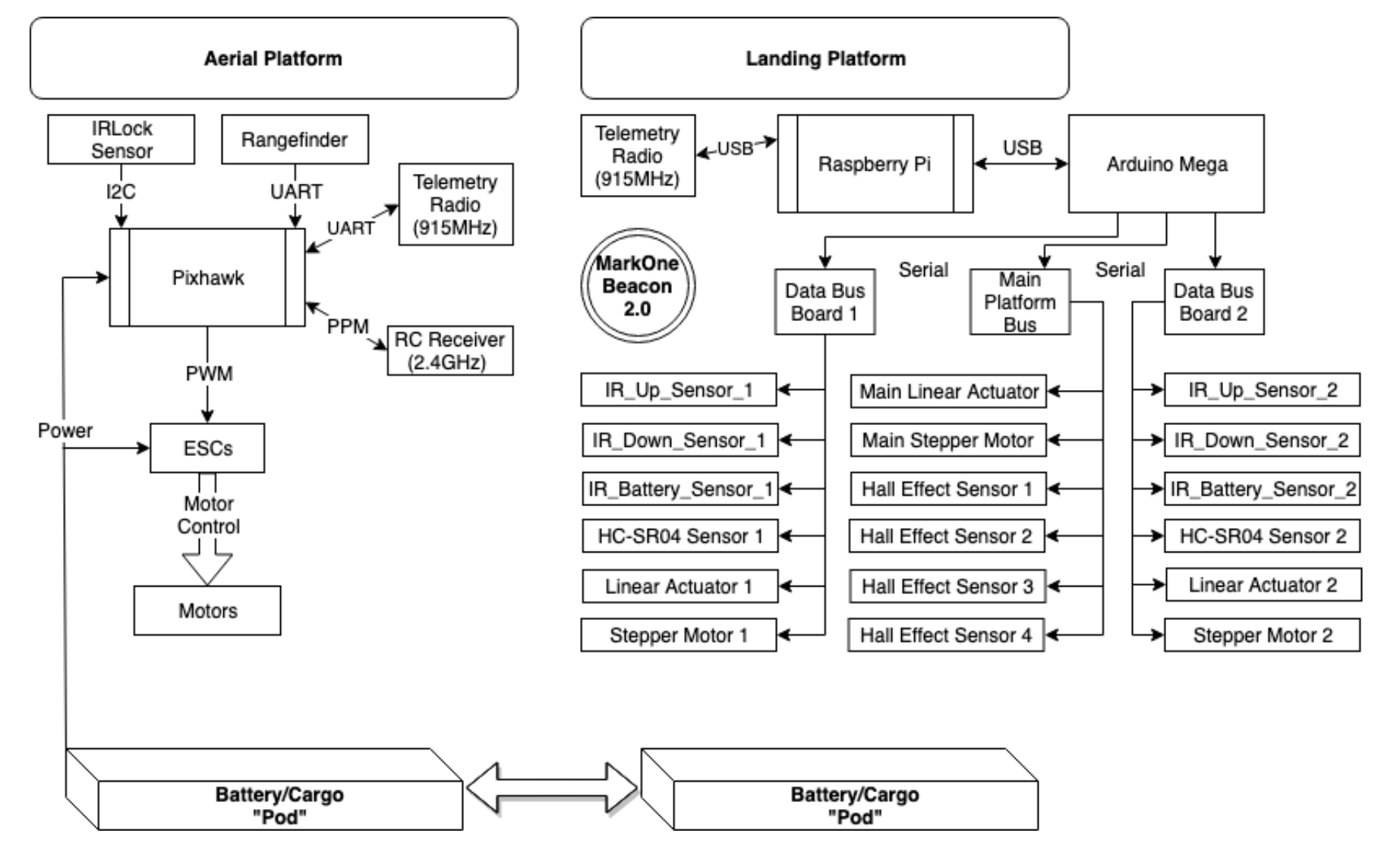}
  \caption{AutoServe System Overview Block Diagram.}
  \label{fig:my_label_9}
\end{figure}

Figure \ref{fig:my_label_9} shows the system architecture of the proposed AutoServe platform, which consists of two main platforms, namely the Aerial Platform (AP) and Landing Platforms (LP). The AP, also known as a drone or UAV, is responsible for carrying cargo, navigating to destination, and landing on the designated LP. The AP is comprised of four main subsystems: the Control System, the ``Pod'' System, the Precision Landing System and the Communication System. The LP is responsible for receiving the incoming APs and performing the scheduled operations for that particular UAV. The LP is comprised of the complement of the the AP's subsystems with the addition of an Alignment System to re-position the UAV into a known orientation after landing.

\subsection{Aerial Platform (AP)}

\paragraph{Flight Control System:} The heart of the AP's control system is a Pixhawk Flight Controller \cite{pixhawk2020pixhawk}. The Pixhawk was chosen due to its extensive documentation with the peripherals required in the design and strong reputation as a reliable option. Between the large array of peripheral connectors and wide-ranging compatibility, this was the best choice for the given price point. The IR-Lock Sensor was connected via I2C, Telemetry Radio via UART (Universal Asynchronous Receiver/Transmitter) and Remote Control Signal via SBUS.

\begin{figure}[htb]
  \centering
  \includegraphics[width=0.5\linewidth]{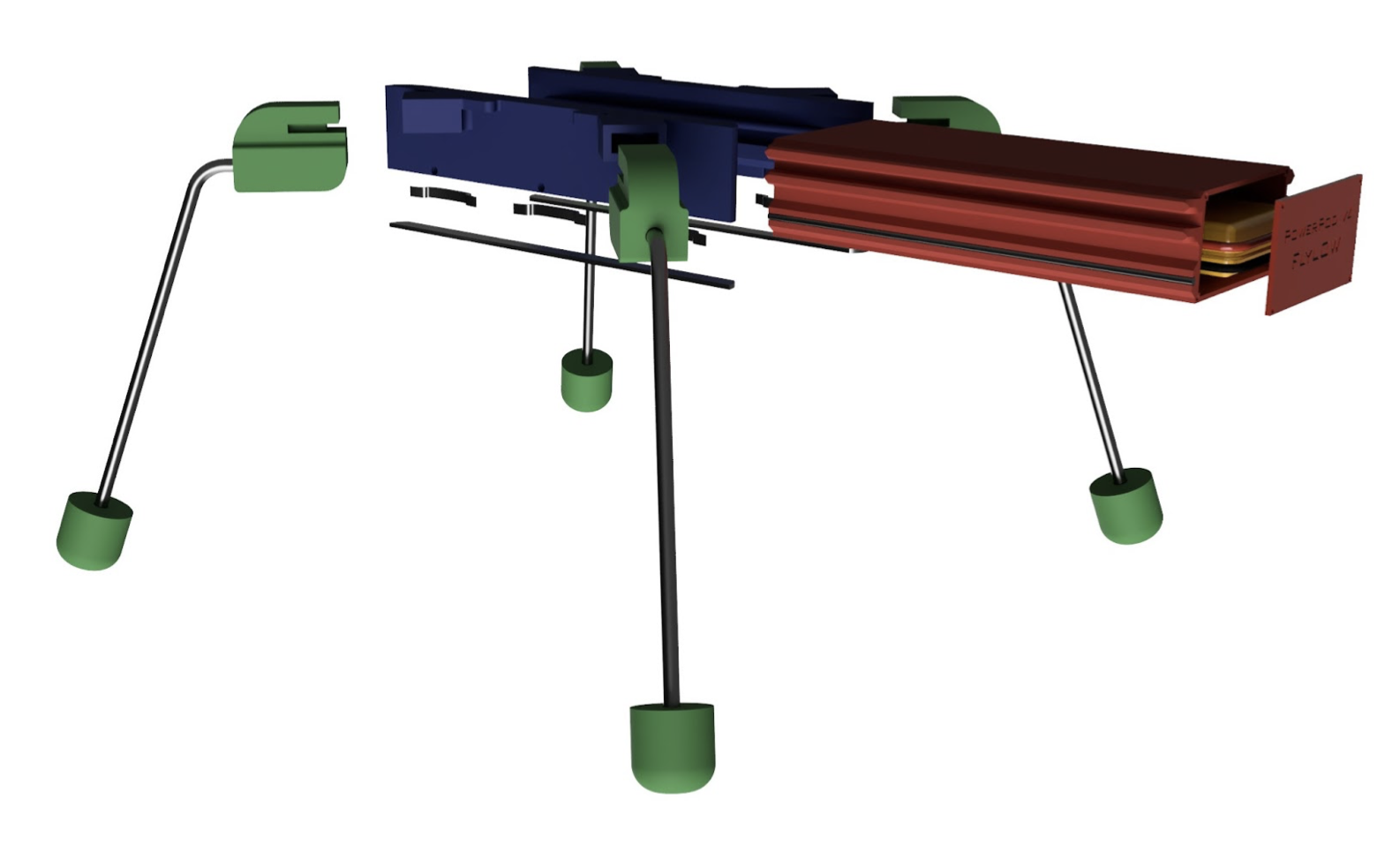}
  \caption{Render of UAV Platform's 3D Printed Parts.}
  \label{fig:my_label_1}
\end{figure}

\paragraph{The ``Pod'' System:} Aiming at a low-cost design, our prototype is built with 3D printing. Using 3D printers to manufacture parts also allowed the team to quickly analyze and iterate through design options. As shown by Fig. \ref{fig:my_label_1}, with polylactic acid (PLA) plastic, we manufactured landing arms, a battery ``pod'' and a rail system for the battery to lock into place and power the UAV. Since the balance connector for the Li-Po battery is only needed for charging, the PowerPod incorporates pads for each balance wire and the rail system only uses BATT+ and GND. The PowerPod acted as a way to standardize the battery pack, paving the way to design a repeatable process of replacing a discharged pod with a fully charged one.

\paragraph{Precision Landing System:} An important aspect of the AutoServe system design is the requirement for the drone to land in a particular orientation and in a relatively small area. Since the GPS-based navigation alone is not sufficient to land within the bounds, the GPS-navigation must be coupled with some other form of position detection. To accomplish this task, the precision landing subsystem's design implemented hardware to ascertain a more accurate position estimate. The first component being a LIDAR rangefinder, which provides the aerial system with an accurate height above ground level (AGL). The other components are an infrared (IR) LED beacon and an IR Camera. Our design made use of the current industry standard, the IR-Lock System which is comprised of the MarkOne IR Beacon and the IR-Lock Sensor. The beacon can be placed on or near the LP, and the UAV will set its landing attitude relative to the location of the beacon's signature in the camera's frame. Each beacon has a unique infrared signature which allows for multiple beacons to be used to identify each LP.

\begin{figure}[t]
  \centering
  \includegraphics[width=0.8\linewidth]{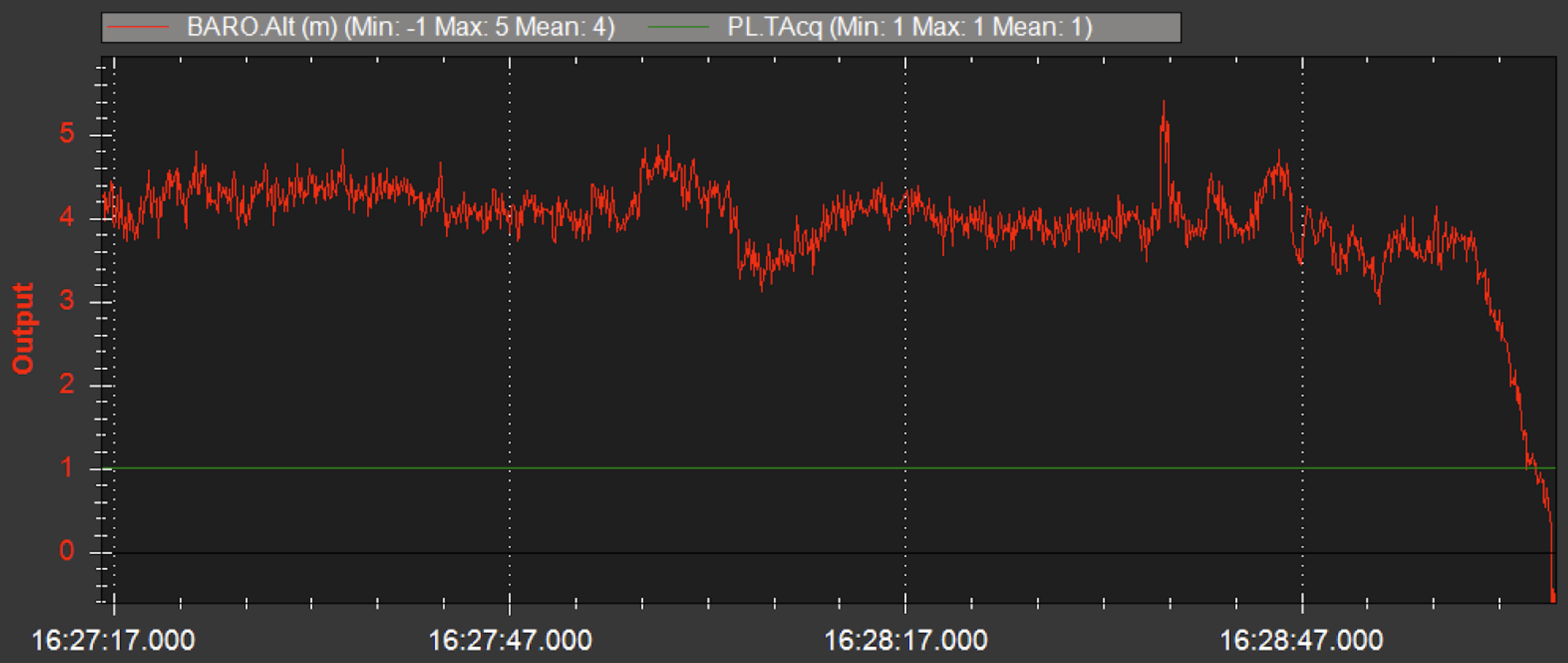}
  \caption{Flight Log with Precision Land Target Acquired (PL.TAcq).}
  \label{fig:my_label_12}
\end{figure}

Through the use of the PX4 Flight Stack \cite{px42020px4}, we were able to use a low-cost LIDAR Rangefinder, the Benewake TFMini. Precision landing requires the use of a rangefinder so the flight controller has a more accurate estimate of the aircraft's height above ground level (AGL). This sensor was connected to the Pixhawk flight controller via UART with a maximum current draw of 100 mA. To use this sensor certain firmware parameters for both the LIDAR and the Pixhawk needed to be configured. TFMini parameters needed to be set through a proprietary software are only available on Windows platforms as programming via serial communication proved unsuccessful. This subsystem enables the UAV platform to land within 10 cm of a stationary target or 30 cm of a target moving at a speed of 1 m/s or less. This minimizes the AP's landing error from meters to centimeters, allowing the design of a compact LP receive, align and maintenance on the UAV. After reviewing flight logs as shown in Fig. \ref{fig:my_label_12}, the operator is able to confirm that the AP was able to acquire the MarkOne Beacon's IR signature and then navigate to land in the designated zone.

\paragraph{Communication System:} In the radio hardware layer, 915 MHz radio links are implemented to support multiple simultaneous MAVLink streams. The chosen telemetry modem was the RFD900+, which was chosen due to its plethora of features at a reasonable price point \cite{RFDesign42020RFDesign}. The most important features of the modem include its 40+ km communication range, point-to-multipoint support and its native integration of the MAVLink communication protocol through the open-source Sik Radio Firmware. With one modem integrated into each platform, telemetry information can be transmitted reliably over large distances. This model is capable of an estimated 250 kbps data rate given ideal transmission conditions.

\subsection{Landing Platform (LP)} 

\begin{figure}[b]
  \centering
  \includegraphics[width=0.5\linewidth]{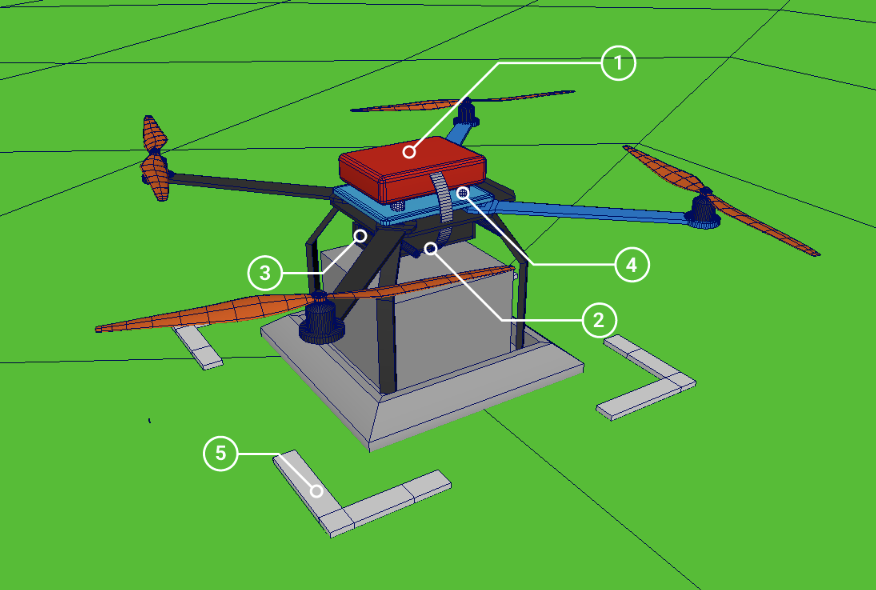}
  \caption{Landing Platform Operation Concept Render: 1: Flight Computer; 2: Infrared Camera; 3: PowerPod; 4: Airframe; 5: Infrared Beacons.}
  \label{fig:my_label_8}
\end{figure}

The LP was implemented with the idea of dispersed \emph{nodes} that were capable of scheduling landings performing maintenance on the incoming APs. Through the use of the precision landing system described above, it is able to land an incoming UAV within a 10 cm radius of the landing beacon. However, in order to simplify the design of the automated battery swapping process it was necessary to align the drone into a known orientation. By use of two sloped linear actuators triggered by the landing, the platform can push the drone into alignment for the necessary maintenance operations. In our case study, the first operation we implemented was automated battery exchange. Figure \ref{fig:my_label_8} shows a landing platform operation concept render, in which all main components are highlighted.

\paragraph{Modular Pod Exchange System:} Once the standardized ``Pods'' had been designed, the next step was implementing a system for repeating the insertion/removal process. This task was accomplished with a linear actuator to push the discharged ``Pods'' out of the AP and replace it with a charged one. The discharged ``Pods'' would be moved into a open charging bay and the charged ``Pods'' would be taken from an occupied one. The ``Pods'' system was conceived purely with the intention of battery swapping but can be expanded further depending on the task to be performed.

\paragraph{Precision Landing System:} The IRLock subsystem acts as the compliment to the AP's Precision Landing System. It is comprised of the MarkOne IR Beacon and is used to identify the LP as well as guide the AP's descent. However, we found the infrared system was not precise enough for our design's overall goal of automated battery swapping. Even with the addition of a Precision Landing system there is still a large enough margin of error to complicate the Pod Exchange system. By adding a separate operation to align the drone after landing, the error of the ``known'' position of the AP is minimized and therefore the Pod Exchange system can be simplified. The alignment procedure is performed by two sequential linear actuators that not only correct rotational error on the landing axis, but also location based error on both the X and Y axes.

\section{Communication Protocol}
\label{sec:comm}

In many distributed systems and networks some resources are shared among multiple entities, such as file system space, application memory, or access to locked data. There is usually a convention to handle both allocation and conflict resolution for the shared resource. In our AutoServe network, the LPs are considered a shared resource when there are more than one UAV in operation. Thus, it is possible that two UAVs request a service from the same LP at roughly the same time. It is also possible that a given LP will not be able to serve an AP quickly enough. In order to deal with dynamically arising conflicts, a clear protocol must be established that is capable of handling these events. This protocol is expected to prioritize APs based on how urgent the task is. For example, how much remaining battery an AP has. It will support dynamic mission states to guide the interactions between different systems. Lastly, it will support a service queue which LPs will maintain to manage their service reservation requests.
    
\subsection{System Health and Priority}

Part of the protocol is to allow each AP to continuously display its system health. Keeping track of these health attributes, such as battery level, will allow the LPs to reserve themselves if required. If an AP is displaying a critical safety value, i.e., requiring a battery swap, then the nearest LP will consider itself reserved, and wait for UAV\_Boarding stage to begin. In the case of battery consumption, ther priority would be a function of the remaining battery. If two UAVs are trying to use the same LP at the same time, the UAV with a lower level of remaining battery will be prioritized. Since addressing conflicts is paramount in our system, the protocol will need to capture some sense of priority among UAVs, similar to handling commercial flight landing requests with one runway.

\subsection{Mission State}

Since the APs and LPs are physical systems interacting with each other, the stages of their cooperation must be passed back and forth. For example, an LP must ensure that an AP has fully landed before starting its alignment system. Because the states of these two systems are dependent on each other, the current state of each system must be contained in the protocol. There also must exist the capability to share the mission state between interacting systems.

\subsection{Service Queue and Reservations}

\begin{figure}[t]
    \centering
    \includegraphics[width=\linewidth]{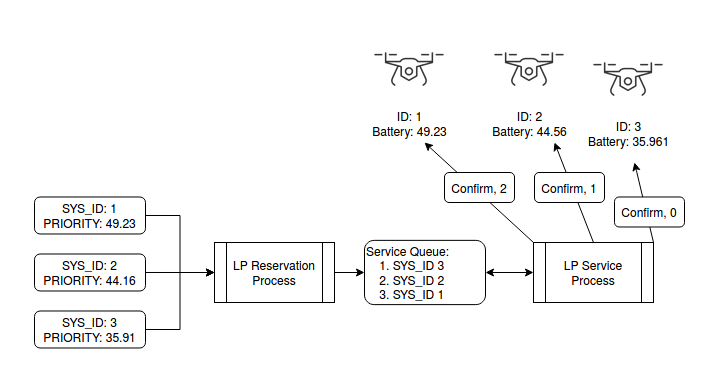}
    \caption{How the LP Handles AP Requests.}
    \label{fig:reservations_and_service}
\end{figure}

In the case that there is more than one AP trying to acquire an LP for service, the APs must be aware of the order that they will be served in. As illustrated by Fig. \ref{fig:reservations_and_service}, to support this AutoServe capability, each LP will store a service queue that signifies the order in which they will service APs. When an AP is popped from the queue, it will then enter its landing phase onto the LP. Once the AP is considered to have fully left the LP, the next AP on the service queue will be chosen.

This service queue is maintained through the concept of reservations. The APs request service reservations from the LP, which will then be responded to with the requesting APs position on the service queue. As discussed previously, it is possible that an LP will not be able to help an AP in time. Thus our protocol's notion of reservations must support keeping and canceling functionality. For example, an AP should be able to request a reservation at a new LP if it cannot be accommodated by the current LP under consideration. 

\subsection{Required Messages}

In order to control the interactions between APs and LPs, the above information needs to be shared in a coherent way. This requires establishing a message set to handle the particular interactions between systems on the AutoServe network. The protocol will support messages for reserving a service slot, maintaining/canceling a reservation, sharing system state, and sharing service queue information. Here are the new messages that our AutoServe network necessitates.

\subsubsection{Service Reservation Request}

The AutoServe system requires some way of signifying that a given AP would like to receive services from a given LP. This takes the form of a reservation request message, which contains two attributes: the AP system priority and the system ID of the LP. Since the network supports more than one LP, it is important for the AP to clarify from which LP it is requesting a reservation. The priority attached to the message will be used to order the service requests.

\subsubsection{LP Reservation Confirmation}

The LP being requested must signify to the requesting AP that its reservation has been received and confirmed. This message requires two attributes: AP system ID and service queue position. Like the previous message, the LP must specify which AP's request it is confirming. It also must give a time estimate, i.e., the queue position, to allow the AP to determine if the LP's reservation accommodates its needs. Once the AP receives a reservation confirmation from an LP with a queue position of $0$, the AP can begin its boarding process.

\subsubsection{AP Reservation Confirmation}

Since the AutoServe network allows APs to reject LP reservation offers when they do not fulfill the AP's needs, the AP needs some way of keeping or canceling the reservation. Along with the ID of the LP recipient, this message body will contain a keep/cancel value that will allow an AP to manage its reservation. In the event that a reservation is canceled, the AP will be removed from the service queue, and all APs with lower priority will be moved up the queue. This message will only be sent from APs to LPs.

\subsubsection{System State}

As stated in the beginning of this section, capturing the concept of system state for APs and LPs is required to facilitate their physical interaction. For example, the LP needs to confirm that the AP is done landing on the platform before changing starting its alignment process. Similarly, the AP must confirm that service is over and that it is no longer restrained before it departs from the LP.

\subsection{The Aerial Perspective}

\begin{figure}[t]
    \centering
    \includegraphics[width=\linewidth]{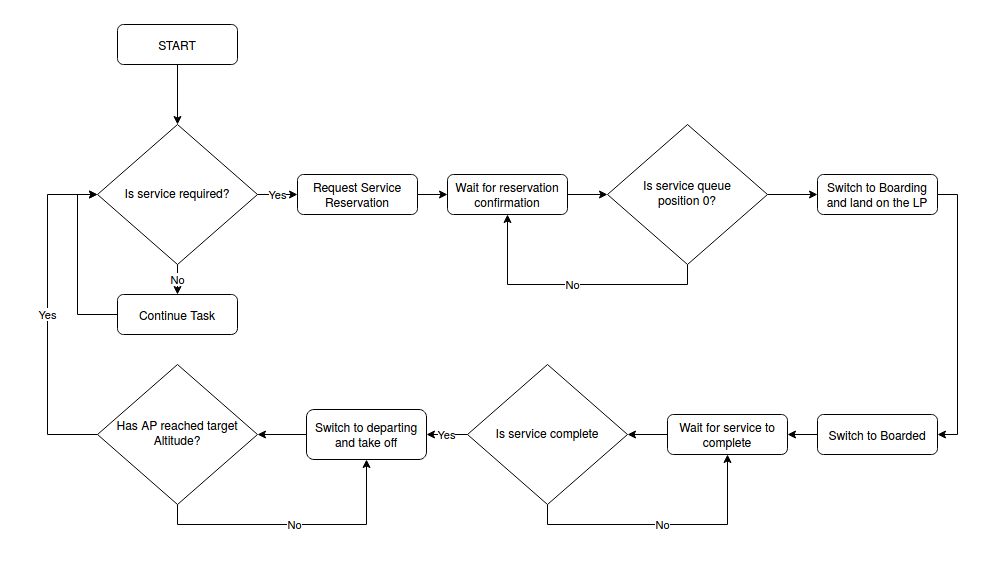}
    \caption{Flow chart: the process of AP Acquires the LP.}
    \label{fig:aerial_perspective}
\end{figure}

When an AP is performing a task and recognizes that it requires services in the near future, it issues a Service Reservation Request to the LP. This request includes a priority, which the LP uses to assign the AP a position on the service queue. Once the AP receives confirmation of the reservation, it will decide if its position on the queue will meet its needs. If the AP needs the service sooner, it can make a service reservation request to a new LP and cancel its reservation with the current LP. 

Once the AP makes a reservation that meets its needs, the AP will then continue its task or perform a ``wait'' maneuver until it gets a service reservation confirmation with its queue position as zero. This is the signal for the AP to begin its boarding process. Once it has boarded, received its service, and taken off, it can continue to complete its required tasks. Figure \ref{fig:aerial_perspective} summarizes the working flow of the process for AP acquires the LP.

\subsection{The Ground Perspective}

Each LP operates in an endless control loop, where it simultaneously adds new reservation requests to its queue and services the given AP with the highest priority. These should be considered as different processes since the LP needs to store new requests while serving a given AP. If the LP finds itself with no APs on its reservation queue, it will respond to the next request with a confirmation and a queue position of 0. This indicates that the AP can begin its boarding procedure.

\begin{figure}[t]
    \centering
    \includegraphics[width=\linewidth]{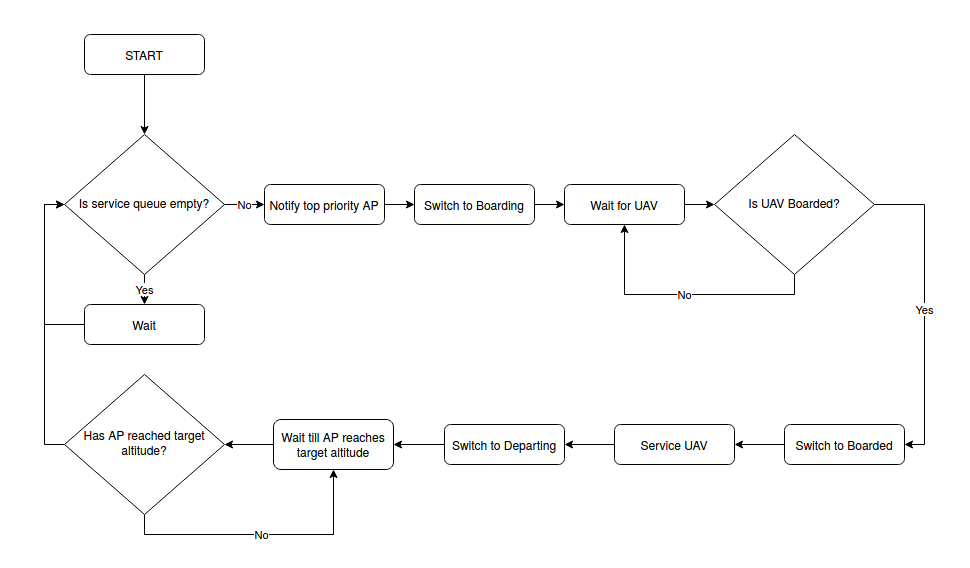}
    \caption{Flow Chart: the process of LP Performs Service on a Given AP.}
    \label{fig:ground_perspective}
\end{figure}

A central part of the LPs protocol design is the priority oriented nature of its reservation queue. Priority queues keep track of a list of data in order of a user defined importance. In the case of an LP, if a given AP requests service before another AP, but the latter has a higher priority, the LP will still process the higher priority request. Figure \ref{fig:ground_perspective} summarizes the working flow of the process of LP Performs Service on a Given AP.

\section{Experimental Study}
\label{sec:exp}

\begin{figure}[t]
  \centering
  \includegraphics[width=\linewidth]{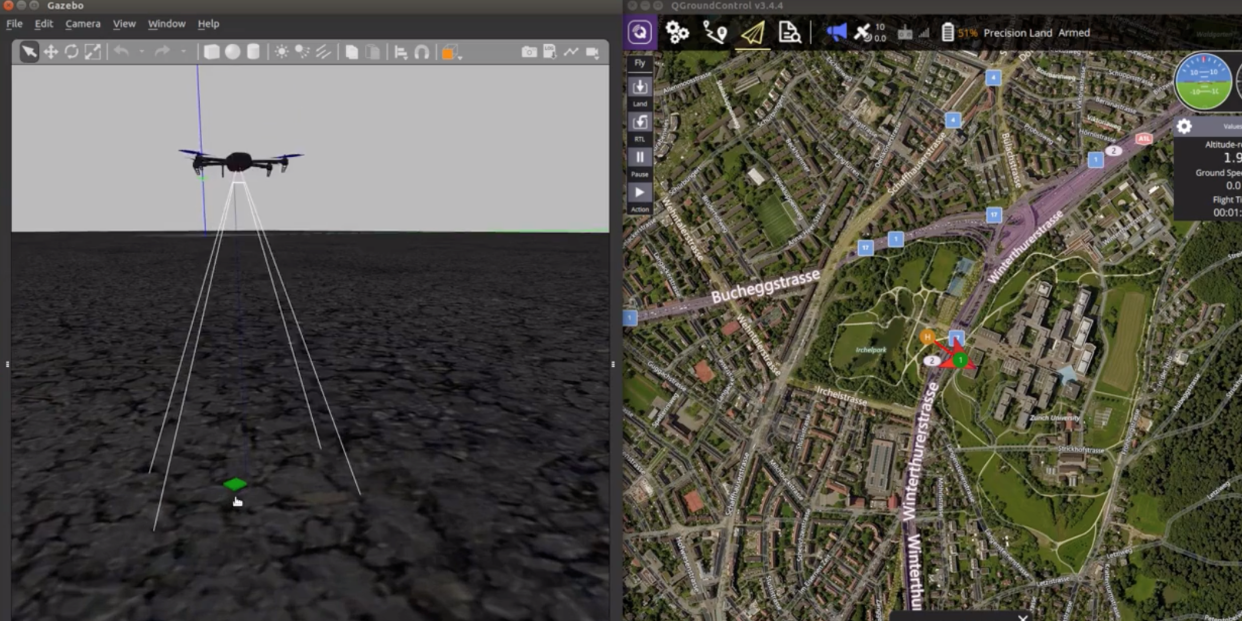}
  \caption{Precision landing simulation using ROS and Gazebo.}
  \label{fig:sitl}
\end{figure}
    
\subsection{Landing Accuracy}

In the realm of UAV software, simulation testing is a crucial step for evaluation. Our software testing was done using ROS's Gazebo simulation environment \cite{koenig_howard_2004}, such as in Fig. \ref{fig:sitl} where we evaluated the ability of our AP to acquire the IR-Lock beacon, and smoothly descend on top of it. In simulation we found the positional error to be within 1 centimeter, even when applying dynamic environmental features like wind. When testing in actual flight, the positional error experienced correlated with previously mentioned error (less than 10 centimeters on a stationary target). This provided enough accuracy to land within the LP's required target.

\subsection{Evaluation of Protocol} 
        
In order to evaluate our protocol we built our own simplified simulation. This simulation tool allows the user to define an amount of UAVs and landing platforms, along with a space for them to exist in. The UAV's ``operate'' in a discrete time manor, where a bounded random amount of battery is consumed, and a random 2D displacement vector is generated to change position. When a UAV notices its battery falls bellow a pre-defined threshold, it will issue a request to the landing platform closest to it. When it services a UAV the operation takes up 120 seconds of operation. Each UAV is also assigned a battery percentage at which it is no longer active. We consider the simulation to fail when the battery level of a UAV on the network drops below this threshold.

For our purposes, we chose to bound our random consumption between 0.15\% to 0.20\% for every second of operation and issues a service request when its battery goes bellow 50\%. The displacement of each UAV was bounded to 0.3 meters per second in either direction. The area of operation for our simulation was 1000 meters by 1000 meters. With these predefined values inside of our simulation, we verified that one LP could accommodate up to five UAVs without any of their battery levels dropping below the threshold of 15\%.

Though these results are promising, there are some significant limitations of our simulation tool that must be considered. The largest limitation is the uncertainty that is associated with software operating in an non-deterministic environment like flying in the real world. Another limitation of this simulation is the resource availability on the landing platform. This means that the LP can service the APs a limitless amount of times. In reality, it would have a limited amount of batteries to replace on the UAVs, or would be limited by the time it takes to charge its used batteries. This simulation is only meant to establish the initial heuristics for an empirical evaluation.
    
\section{Conclusions and On-Going Efforts}
\label{sec:conclu}

The limitations on a UAV's ability to independently perform a function are the main restrictions on its practical application. The field has achieved many of the necessary in-flight requirements for self sufficiency, but has not fully addressed any of the tasks required to keep a system flying. By addressing the moments that need human oversight with a physical service network for drones paired with a custom protocol, in this paper we introduce an autonomous service network infrastructure (AuroServe), which brings a new level of autonomy to UAVs and dramatically increase the domain of their practical application. 

This paper reports our design rationale and some preliminary results. The on-going efforts will address multiple aspects of the system:

\begin{itemize}
    \item \emph{Further Evaluation of Protocol}: Since the occurrences of the lazily evaluated requests are difficult to predict, a more comprehensive study must be performed to determine the true overhead of a custom message.
    
    \item \emph{Mobile Landing Platform}: While the original concept was designed with stationary LPs in mind, this can be expanded upon greatly. A vehicle mounted landing platform has many applications with major benefits to both cost and operational efficiency.
    
    \item \emph{Adding Platform Services}: Battery charging and replacement is only one aspect possible supportive activities. Some next steps that address modern challenges could be adding services for propeller replacement, cargo management, calibration, sheltering from weather, etc.
    
    \item \emph{Security Upgrades}
          \begin{itemize}
              \item \emph{Software MAVLink Message Encryption}: Since MAVSec has only been tested in simulation, further study into low-resource encryption is needed because securing these messages is required to operate safely while maintaining an efficient communications link between UAVs and GCSs.
              
              \item \emph{Hardware MAVLink Message Encryption}: By separating the command and control channel from other higher bandwidth services, the system's designer can implement an independent radio channel for long-range and low-resource communication, which in this case would be MAVLink messages. More expensive telemetry radios, such as the RFDesign RFD900x Modem, integrate hardware accelerated 128-bit AES encryption of messages. This added layer of security allows for command and control messages to be encrypted without any added work being performed by the flight computer.
          \end{itemize}
          
    \item \emph{Extending to Other Autonomous Vehicles}: The automotive industry has been making advances towards battery powered vehicles autonomous vehicles. As the vehicles change, so will the infrastructure we design for them. A very similar battery replacement network could be provided for autonomous cars in order to avoid long charging times and increase system autonomy.
\end{itemize}



\bibliography{report}   
\bibliographystyle{spiejour}   

\end{document}